\documentclass[sigchi]{acmart}
\def\BibTeX{{\rm B\kern-.05em{\sc i\kern-.025em b}\kern-.08emT\kern-.1667em\lower.7ex\hbox{E}\kern-.125emX}}

\copyrightyear{2019} 
\acmYear{2019} 

\title[Multitasking with Alexa]{Multitasking with Alexa: How Using Intelligent Personal Assistants Impacts Language-based Primary Task Performance}

\usepackage{booktabs} 
\usepackage{ccicons}  
\usepackage[utf8]{inputenc}
\usepackage{natbib}
\usepackage{hyperref}
\usepackage[justification=centering]{caption}
\usepackage{titlesec}
\usepackage{subcaption}

\copyrightyear{2019}
\acmYear{2019}
\acmConference[CUI 2019]{1st International Conference on Conversational User Interfaces}{August 22--23, 2019}{Dublin, Ireland}
\acmBooktitle{1st International Conference on Conversational User Interfaces (CUI 2019), August 22--23, 2019, Dublin, Ireland}
\acmPrice{15.00}
\acmDOI{10.1145/3342775.3342785}
\acmISBN{978-1-4503-7187-2/19/08}

\begin{document}

\author{Justin Edwards*}
\author{He Liu**}
\author{Tianyu Zhou**}
\affiliation{%
  \institution{University College Dublin}
  \city{Dublin}
  \country{Ireland}
}
\email{* first.last@ucdconnect.ie}
\email{** first.last1@ucdconnect.ie}

\author{Sandy J. J. Gould}
\affiliation{%
    \institution{University of Birmingham}
    \city{Birmingham}
    \country{United Kingdom}
}
\email{s.gould@cs.bham.ac.uk}

\author{Leigh Clark***}
\author{Philip Doyle**}
\author{Benjamin R. Cowan***}
\affiliation{%
  \institution{University College Dublin}
  \city{Dublin}
  \country{Ireland}
}
\email{*** first.last@ucd.ie}

\renewcommand{\shortauthors}{J. Edwards et al.}

%
%

\begin{abstract}
Intelligent personal assistants (IPAs) are supposed to help us multitask. Yet the impact of IPA use on multitasking is not clearly quantified, particularly in situations where primary tasks are also language based. Using a dual task paradigm, our study observes how IPA interactions impact two different types of writing primary tasks; copying and generating content. We found writing tasks that involve content generation, which are more cognitively demanding and share more of the resources needed for IPA use, are significantly more disrupted by IPA interaction than less demanding tasks such as copying content. We discuss how theories of cognitive resources, including multiple resource theory and working memory, explain these results. We also outline the need for future work how interruption length and relevance may impact primary task performance as well as the need to identify effects of interruption timing in user and IPA led interruptions.
\end{abstract}


\copyrightyear{2019} 
\acmYear{2019} 
\acmConference[CUI 2019]{1st International Conference on Conversational User Interfaces}{August 22-23, 2019}{Dublin, Ireland}

\maketitle

\titlespacing*{\section}{0pt}{0.5ex plus 1ex minus .2ex}{0.5ex plus .2ex}
\titlespacing*{\subsection}{0pt}{0.5ex plus 1ex minus .2ex}{0.5ex plus .2ex}

\section{INTRODUCTION}
Intelligent personal assistants (IPAs) like Amazon Alexa and Google Assistant are now commonplace and easily accessible through devices such as smartphones and smart speakers. Users find these IPAs useful for multitasking in hands-busy, eyes-busy situations \cite{aylett_none_2014, cowan_what_2017, luger_like_2016}. They allow users to complete tasks simultaneously, such as gaining information or launching applications whilst writing. Although multitasking is perceived as a primary benefit of IPA use, it is unclear what the impact of using these assistants is on primary task performance. This impact is especially interesting for scenarios in which peoples’ primary tasks involve similar mental resources to IPA use, like speaking or writing, where language is also being generated. As IPA use grows, it is vital that we understand the impact of multitasking with these devices on user performance. We take a step towards this by exploring in particular how writing tasks are impacted by IPA interaction. Through a controlled experiment, we find that IPA interruptions were significantly more disruptive to writing tasks when the writing task was cognitively demanding and involved significant levels of language generation, utilizing similar resource capacity to IPA use. The work highlights the importance of researching IPA-based interruptions, identifying how overlap in cognitive resources may have a significant role to play in IPA multitasking effectiveness. 

The work presented looks to focus specifically on the impact of speech-based interruptions on primary tasks that we may conduct that also use language. In this research we explore tasks whereby users may be copying text (e.g. transcribing notes) or synthesizing text to generate new text content (e.g creating summaries of text) as a primary task. We hypothesise that synthesising tasks will lead to a higher cognitive load than copying tasks (H1). Based on multiple resource theory and the higher cognitive load for synthesising content, we hypothesize that interruptions using IPAs will be more detrimental to performance when synthesising rather than copying text based content (H2). By demonstrating the extent to which effects of shared cognitive resources between tasks affect spoken computer interactions, like in IPA use, we hope to take steps toward application of existing multitasking and interruptions theory into IPA design and use case considerations. Furthermore, by using IPAs in a dual-task experiment like those prominent in other multitasking work \cite{iqbal2010cars, mark_pace_2012}, we hope to explore a methodology by which researchers can better understand IPA multitasking behaviour.

\section{RELATED WORK}
Multitasking has been investigated in numerous human-computer interaction (HCI) contexts, in particular driving \cite{iqbal2010cars} and the workplace \cite{mark_pace_2012}. Multitasking and interruptions are seen as a singular phenomenon on a continuum of time between task switches, from concurrent to sequential multitasking, with attention staying on a given task on the order of a few hundred milliseconds to a few hours \cite{salvucci2009toward}. Previous research on the impact of interruptions has shown that they are generally detrimental to primary task performance \cite{bailey_effects_2001}, particularly in situations where people are already experiencing high cognitive load \cite{bailey_effects_2001, iqbal_leveraging_2006}.

 Users perceive IPAs as useful tools to help with multitasking \cite{luger_like_2016,cowan_what_2017}. Yet current work on IPAs has focused more on user experience issues \cite{luger_like_2016, cowan_what_2017, dubiel_survey_2018, porcheron_talking_2017}, than examining the impact of task performance when multitasking in particular. This work highlights that people tend to use IPAs to conduct simple tasks such as asking factual or trivia questions, playing music, getting weather updates and setting alarms and reminders  \cite{luger_like_2016, dubiel_survey_2018}. Interaction tends to be brief, often taking the form of isolated question/answer dialogues \cite{porcheron_voice_2018}. Although some users interact frequently with IPAs, they tend to mistrust their ability to execute tasks effectively, especially in more complex multi-turn tasks \cite{luger_like_2016}. Indeed users have also questioned how supportive current IPA design is to hands free use \cite{cowan_what_2017}. Nevertheless, users express a desire for IPAs to support their other tasks in a seamless, context-aware way \cite{luger_like_2016} that is in line with the description of these systems as \textit{intelligent} personal assistants. 
 
Although little work has specifically focused on IPA multitasking, there has been research on how listening to speech during dual-task contexts affects performance. This has focused mainly on the influence of listening to noisy speech while performing a tactile \cite{fraser2010evaluating} or visual secondary task \cite{desjardins2014effect}. Noisy speech is a major source of increased cognitive load in these listening tasks, reducing performance on other tasks. Listening to synthesized speech, the type of speech delivered by IPAs,  while performing a visual task also seems to impact primary task performance. Listening to high-quality synthesized speech has been shown to lead to better visual task performance than when listening to low-quality synthesized speech or even natural human speech \cite{govender_measuring_2018}. Our work seeks to go beyond merely listening to speech, looking instead at IPA interactions which involve both speech production and listening to synthesized speech \cite{porcheron_voice_2018}, a popular interaction paradigm that has not been investigated in this way. Like previous dual-task studies involving synthesized speech \cite{fraser2010evaluating}, our work presents tasks in different modalitlies. While this methodological decision reduces overlap between cognitive resources \cite{wickens_multiple_2002} and resembles realistic IPA use cases \cite{luger_like_2016}, we expected our work would follow previous multimodal dual-task experiments in nonetheless detecting difference in multitasking and monotasking performance \cite{fraser2010evaluating, Alsius2007}.  

In terms of IPA use, interruptions literature supports the proposition that primary task interruption with an IPA may be detrimental to primary task performance \cite{bailey_effects_2001}. Interruptions involving IPAs may be especially impactful when the interrupting (or secondary task) and the primary task both share the same resources needed for task execution. In an IPA context, this may be when conducting primary tasks that involve language processing and production (such as writing), which are also used to generate IPA commands. This prediction is supported by multiple resource theory \cite{wickens_multiple_2002}. The theory proposes that cognitive resources are separated across numerous dimensions including modality, code, and stage. \emph{Modality} is described as the perceptual channel by which information is processed, including visual and auditory channels. For IPA interactions, this would typically be auditory, but it may be multimodal (i.e. auditory and visual) for devices that use screens to support IPA use (e.g. Siri, Google Assistant). \emph{Code} defines the distinction between spatial information and symbolic or verbal processes. In multitasking literature, this distinction is relevant to describing differences between the processes being considered in tasks e.g. differentiating visually tracking a moving object from reading or listening to speech. For making IPA requests, code would always be verbal due to the fact that speech is a verbal process of interaction. \emph{Stage} defines the distinction between 1) perception and 2) the planning/execution of user responses \cite{wickens_multiple_2002}. A task like IPA interaction would include both of these stages as an interlocutor needs to hear and perceive their partner’s speech as well as plan and produce their own in response. For IPA interaction, the user starts in the execution stage as interactions are always initiated by the user speaking first, meaning perception does not begin until after the IPA has made an utterance. 

Multiple resources theory holds that concurrent tasks are more disruptive to one another when they have higher overlap across these dimensions. This has been demonstrated experimentally as auditory interruptions are less disruptive than visual interruptions to visual primary tasks \cite{ratwani_effect_2008, ho_not_2004}. Similar cognitive resources to those used in writing new content (e.g. \cite{grabowski_premotor_1998})  may also be used when generating IPA queries as both tasks involve planning and production of language, meaning potentially significant disruption to such primary tasks. This may be especially true if the primary task is also of high cognitive load. Demonstrating this would indicate that writing tasks or tasks that overlap in resources with querying IPAs are particularly impacted by IPA use and perhaps particularly ill-suited for IPA multitasking, a valuable insight to designers and users. 
 
\section{METHOD}
\subsection{Participants}
Twenty-four students from a European university (M=12, F=12; mean age=25yrs, SD=5yrs) took part in the study. Participants were recruited via emails sent to their school mailing lists and flyers posted on campus. As thanks for taking part participants were included in a prize draw for three €25 vouchers. All participants reported being native or near-native English speakers with the majority being Irish (46\%). Ratings of self reported typing and voice interface experience (7-point Likert scale:1=Not at all experienced - 7=Very experienced) suggested that the sample on average had high levels of typing (M=5.04; SD=1.45) and low levels of voice based interfaces experience (M=2.75; SD=1.78). 

\subsection{Primary Task}
Participants completed two types of typing task whereby they had to synthesize and generate (\textit{rewording  task condition}) or copy word-for-word paragraphs of text (\textit{copying task condition}) in a within-participants design. The texts used in the experiment materials were derived from educational texts from a number of language learning sites \cite{joyen37, hujiangship, listeningexpress} to ensure participants would find the material accessible and easy to understand. All texts were of similar length (between 980 and 1150 characters; between 185 and 188 words). These tasks were chosen as they represented realistic typing tasks but were simple enough that participants would not struggle with completing them.

\subsubsection{Rewording Task Condition:}
In the rewording task condition, participants were asked to reword a paragraph of text. They were instructed to make sure that they preserved the essence and meaning of the paragraph in their rewording. Rewording text involves a significant number of cognitive processes (e.g. reading, processing and comprehending the existing text, holding this text in memory whilst synthesising, planning and translating ideas into language, see \cite{mccutchen_capacity_1996, foroughi_interruptions_2015}), making it more likely that it require significantly higher cognitive load compared to the copying task (see below). 

\subsubsection{Copying Task Condition:}
When in the copying task condition, participants were asked to copy word for word the paragraph of text displayed. This task is assumed to cause lower cognitive load as, unlike the rewording task, it does not require deep processing about the meaning of the text or any planning of how to reiterate that meaning. In other words, synthesis of the text, planning and translation are significantly reduced in this condition as the task only requires direct copying of the sentences in the excerpt.

\subsection{Secondary Task}
While engaging in primary tasks participants were asked to complete two secondary tasks with an Amazon Echo Dot (\textit{Interruption trials}). There were also trials in which participants were not asked to interact with the IPA while engaging in the primary task (\textit{No Interruption trials}). The No Interruption trials were included to ensure that participants were not able to predict the sessions in which they experienced interruptions. Because the research focuses on the impact of interruptions on user task performance, the No Interruptions trials were not included in the task performance analysis. This is standard in multitasking literature, whereby behavioral effects of interruptions (i.e. lags and errors) are only compared for conditions that actually include interruptions  \cite{gould_what_2013, brumby2013recovering}. 

\subsubsection{IPA Requests:}
When in the Interruption trials, participants were asked to interact with the IPA to get the answers for 2 out of a possible 6 questions. The questions reflect common tasks and requests made to IPAs \cite{luger_like_2016, cowan_what_2017, dubiel_survey_2018, porcheron_talking_2017}. These included: \begin{itshape}1) What’s the weather like in Dublin?; 2) What is the tallest building in the world?; 3) Who is the fastest runner in the world? ; 4) what’s the tallest mountain in the world?; 5) What does IELTS stand for?; 6) What is the longest river in the world?\end{itshape}  The cognitive processes involved in user’s generating speech requests to answer these queries in the Interruption trials, are assumed to share similar resources as the translating part of writing (e.g. \cite{grabowski_premotor_1998}), suggesting a significant overlap of stage when performing the rewording primary task.

The questions were randomized across the experiment. All IPA responses were approximately 11 seconds in duration. Once the IPAs response was finished participants were free to return to the primary task. The interruption of participant’s primary task occurred twice within each primary task condition. The points in the paragraph where participants were to be interrupted were randomly determined before the study and were consistent for all participants. Participants were alerted when to make a request to the IPA through a notification on the screen, sent by a member of the experiment team who was in another room remotely viewing the participants screen to monitor their progress. When participants reached the point in the primary task where the interruption was to be delivered they were sent the on screen message, with a question to answer using the IPA. Once the participant had made the speech request the remote experimenter then played pre-prepared audio to answer the query through the Amazon Echo Dot placed next to the user. This therefore simulated the use of the Echo Dot in the experiment whilst controlling for potential effects of misrecognition or errors in the Echo determining the user query. The audio used for the responses delivered by the IPA were produced using Google WaveNet text-to-speech synthesis \cite{oord_wavenet:_2016}.

\subsection{Measures} 
\subsubsection{NASA TLX Score:}
To test whether the primary tasks varied in cognitive load, we asked participants to complete a NASA-TLX questionnaire \cite{hart_development_1988} after each completion of the primary tasks. The NASA-TLX is a 6-item Likert scale (20 point scale per item) questionnaire, measuring 6 concepts related to mental workload; \begin{itshape}Mental Demand, Physical Demand, Temporal Demand, Performance, Effort and Frustration\end{itshape}. It has been widely used in a number of domains to assess workload, ranging from HCI user-evaluations of prototype devices \cite{cockburn2003evaluating} to industrial psychology studies of medical professionals \cite{tubbs2013observational}. Mental workload has been demonstrated to be a contributing factor to cognitive load \cite{GALY2012269}. The NASA-TLX measurement of mental workload is an easily administered tool that is sensitive to changes in intrinsic cognitive load \cite{wiebe2010examination}. Scores on the questionnaire can be summed to create a Raw TLX score used to identify overall workload  (see \citet{hart_nasa-task_2006} for discussion).

\subsubsection{Interruption Lag:}
As operationalised in \citet{trafton_preparing_2003}, interruption lag was measured as the time difference (in hundredths of a second) between the delivery of the notification of the secondary task to the user and the start of the participant’s utterance to the IPA. Screen recordings were analyzed to measure when the interruption notification appeared on a participant’s screen and when the participant began their utterance. 

\subsubsection{Resumption Lag:}
Again, following \citet{trafton_preparing_2003} resumption lag was measured as the time difference between the IPA finishing the delivery of the information and the resumption of typing on the primary task. Screen recordings were analyzed to measure when the IPA concluded its utterance and when the participant resumed typing for the primary task.

\subsection{Procedure}
Participants were recruited via email and through snowball sampling. When arriving at the lab participants were given information about the study and asked to give informed consent before taking part. Participants then completed a demographics questionnaire collecting data about their age, sex, educational background, their experience with voice based technologies and their typing skill. Further details about the writing tasks were then given, with participants being told 1) they would need to complete four writing tasks; 2) that they will be typing words into a blank word processing document; 3) that two of these tasks would involve copying a short piece of text 'word-for-word'; 4) that the other two tasks would involve rewording the text shown; and finally, 5) that they would have ten minutes to complete each writing task. Task order was randomly assigned to mitigate order effects. Before starting, participants were also informed that whilst doing the writing tasks they may be asked to request some information from the Amazon Echo Dot placed on the desk, beside the computer they were using for the writing task. They were told that the information they needed to ask the Echo for would be shown in a window at the bottom of the screen. Participants were told to use the Echo Dot when this information appeared to request this information using their voice and return to the writing task once the Echo had finished giving the information. Participants were told to complete the tasks as quickly and as accurately as possible.

So as to familiarise the participants with the secondary task delivery process and how the Echo Dot responded, they were given two practice tasks (one copying and one rewording task). In one of these they did not experience an interruption and in one they did experience an interruption. Experiment sessions were recorded using Flashback Express, a screen recording software package. This captured the user screen interaction and the audio of the session. Two experimenters reviewed videos of the sessions to extract timing data used in the lag analyses below. After each primary task, participants then completed the NASA-TLX questionnaire, resulting in four completed NASA-TLX questionnaires per participant. Upon finishing the experiment, participants were thanked for taking part and debriefed as to the motivations of the study.

\begin{figure}
  \includegraphics[width=0.4\textwidth]{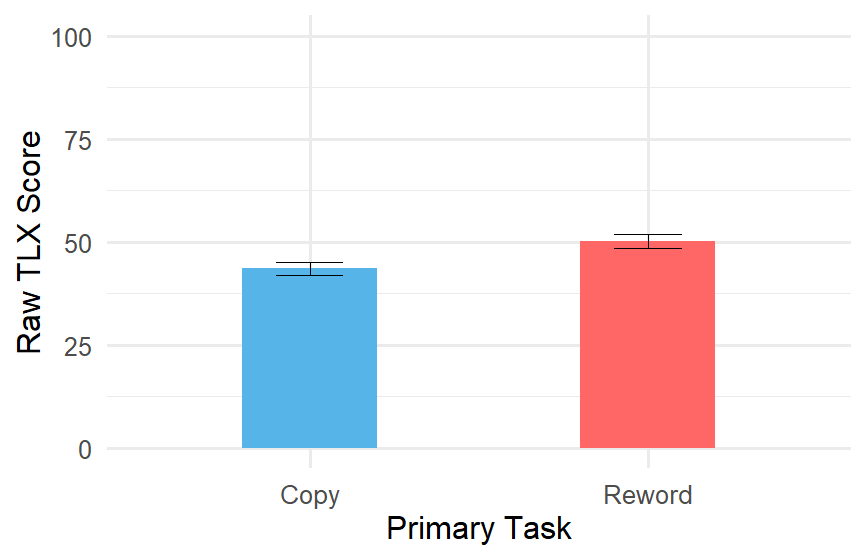}
  \caption{Mean Raw TLX score (with standard error) for each primary task condition}
  \label{fig:tlx}
\end{figure}

\section{RESULTS}
\subsection{Raw TLX Scores:}
A one-way paired samples t-test was run to analyse the effect of the primary task conditions on subjective workload. We found that, as predicted, there was a statistically significant effect of primary task type on subjective workload ratings, t(47)=-3.12, p=.002, d=0.45, (see Figure \ref{fig:tlx}). Participants rated the rewording task (M=50.21, S.D.=15.71) significantly higher in terms of mental workload than the copy task (M=43.60, S.D.=15.14), supporting H1.

\subsection{Lag Analysis:}
To identify the effect of interruptions of the primary tasks, lag times from the Interruption condition only were analysed. This is because there were no interruptions in the No Interruption condition and thus measurement of interruption and resumption lag were not possible. Linear mixed effects models were run using the \emph{lme4} package (Version 1.1-7) \cite{bates2012package} in R (Version 3.5.1) \cite{rcore} to analyse the effect of interruptions on the interruption and resumption lag. Linear mixed effects models are an extension of linear regression. Like linear regression they allow us to model fixed effects, yet they also allow us to model co-dependency - participant and material based variation within the data - through random effects (see \citet{barr_random_2013,baayen2008mixed,baayen2010analyzing,singmann_introduction_nodate} for discussion). This leads to more generalisable effects and improved statistical power as per-participant and per-task effects are mitigated \cite{singmann_introduction_nodate}. Both models for the interruption and resumption lag included within-participant random slopes for primary task type.

\subsubsection{Interruption Lag:}
\begin{figure}
  \includegraphics[width=0.45\textwidth]{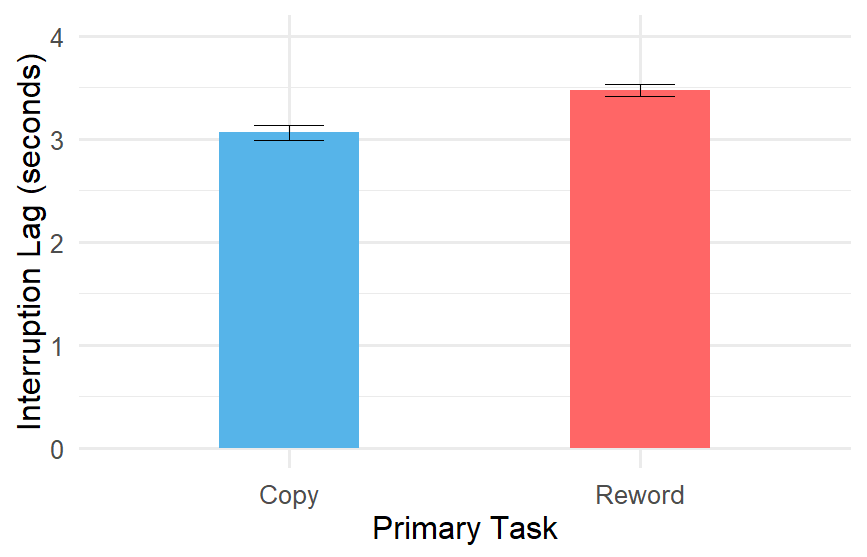}
  \caption{Mean interruption lag (with standard error) for each primary task condition}
  \label{fig:interrupt}
\end{figure}

The mixed effects model showed that there was a statistically significant fixed effect of primary task type on interruption lag (Unstandardized $\beta$=0.41, SE $\beta$= 0.11, p=.003). Interruption lag was significantly higher in the rewording task condition (M=3.47s, SD=0.96) than in the copying task condition (M=3.07s, SD=0.63). The effect is shown graphically in Figure \ref{fig:interrupt}. This supports our hypothesis (H2). 

\begin{table}[]
\caption{Summary of fixed and random effects for interruption lag LME model \\\hspace{\textwidth} Model: Interruption Lag\textasciitilde Task + (1+Task|Participant)}
\begin{tabular}{lllll}
Fixed Effects  & Unstandardized $\beta$ & SE $\beta$   & t  & \emph{p}        \\ \hline
Intercept      & 3.07      & 0.12 & 26.25   & \textless{}.001 \\
Task (Reword)  & 0.41      & 0.11 & 3.99    & \textless{}.001 \\ \hline
               &           &      &         &                 \\ \hline
Random effects &           &      & SD      & \emph{r}           \\ \hline
Participant    &           &      &         &                 \\
       & Intercept           &      & 0.54    &                 \\
  & Task (Reword)          &      & 0.44    & 0.80           
\end{tabular}
\end{table}

\subsubsection{Resumption Lag:}
\begin{table}[]
\caption{Summary of fixed and random effects for resumption lag LME model \\\hspace{\textwidth} Model: Resumption Lag\textasciitilde Task + (1+Task|Participant)}
\begin{tabular}{lllll}
Fixed Effects  & Unstandardized $\beta$ & SE $\beta$       & t  & \emph{p}         \\ \hline
Intercept      & 3.64      & 0.27 & 12.56   & \textless{}.001 \\
Task (Reword)  & 1.91      & 0.32     & 5.91    & \textless{}.001 \\ \hline
               &           &          &         &                 \\ \hline
Random effects &           &          & SD      & \emph{r}          \\ \hline
Participant    &           &          &         &                 \\
       &      Intercept       &          & 1.22    &                 \\
     &    Task (Reword)       &          & 1.43    & 0.96           
\end{tabular}
\end{table}

\begin{figure}
  \includegraphics[width=0.45\textwidth]{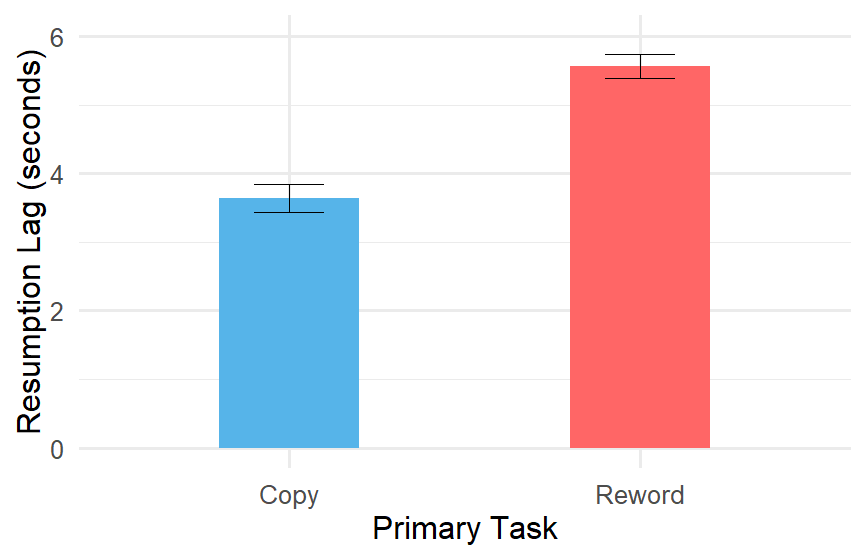}
  \caption{Mean resumption lag (with standard error) for each primary task condition}
  \label{fig:resumption}
\end{figure}

The mixed effects model showed that there was a statistically significant fixed effect of primary task on resumption lag  (Unstandardized $\beta$=1.91, SE $\beta$= 0.34, p<.001). There was a longer resumption lag when participants completed the rewording task (M=5.56, S.D.=2.72) than when they were completing the copying task (M=3.64, S.D.=1.46). The effect is shown graphically in Figure \ref{fig:resumption}. This also supports our hypothesis (H2).

\section{DISCUSSION}
Our work looked to identify the role that interruptions using speech interfaces such as IPAs have on performance of primary tasks, in particular primary tasks that also involve the production of language. We found that multitasking using an IPA when people have to complete tasks that involve generating language led to significantly higher interruption and resumption time costs when compared to writing tasks where the language had to be directly copied. Below we explore potential reasons for this effect and our planned future work in this area.

\subsection{Cognitive resource use in IPA multitasking}
 As noted, both the interruption lag and resumption lag associated with an IPA speech task were greater with a primary rewording task than with a primary copying task, as hypothesized. Echoing previous research on interruptions \cite{iqbal_leveraging_2006,borst_problem_2010}, we found that the primary task that led to more cognitive load as mediated by differences in mental workload (e.g. the rewording task) was more disrupted by IPA use than one that had lower perceived load (e.g. copying task). These findings can be framed in terms of multiple resource theory \cite{wickens_multiple_2002}, where the similarity of resources being used across primary and secondary tasks give can explain the effects seen. While both primary tasks shared a \begin{itshape}{code} \end{itshape} (verbal) with the interrupting task and neither shared its \begin{itshape}{modality} \end{itshape} (speech), rewording is more similar to the secondary task in the dimension of \begin{itshape}{stage} \end{itshape}. Copying requires very little planning in terms of strategy selection as participants produce exactly the same words that they have read, eliminating the need to decide on which words to use. The rewording task and requesting information from IPAs, on the other hand, both require the participant to plan their word selection before generating language. Though participants were supplied a request in the secondary task, they nevertheless had to decide how to structure their utterance. This overlap in resources compared to the copying tasks may explain why this task in particular led to greater disruptive effects. 

It is important to note that the differences in load on working memory components \cite{baddeley1974working}, in particular the phonological loop, across the primary tasks may also explain these effects. Participants likely needed to hold more information in working memory for longer while rewording than copying. As planning an utterance would add verbal information to working memory, the IPA interruptions may have made returning to the rewording task more dependent on working memory capacity.

Future studies could look to identify whether working memory capacity or the overlap between multiple resources play a larger role in IPA interruption impact. In terms of multiple resources, work could look to isolate the effect of \begin{itshape}{code}\end{itshape}, \begin{itshape}{modality} \end{itshape}and \begin{itshape}{stage}\end{itshape} to identify the potential role each of these has on these interruptions. For instance, \begin{itshape}{modality}\end{itshape} could be isolated by selecting two primary tasks that are identical but presented in different modalities, like presenting the reword task from this study but varying whether participants respond by typing or by speaking. Alternatively, \begin{itshape}
{code} \end{itshape} could be isolated by presenting tasks that differ only in whether they use spatial or symbolic stimuli. This could be similar to the copy task from the present study as compared against a task in which participants copied shapes or pictures. Further work on the cognitive mechanisms in type of multitasking would add theory based insight into what sorts of tasks speech interfaces are most appropriate and beneficial in supporting. This would also answer calls in the field for more theory-based understanding of speech interface interactions more generally \cite{clark_state_2018}.

\subsection{Task Relevance and Interruption Length}
 Because interruptions in this work were always irrelevant to the primary task, all interruptions required participants to keep in memory a model of their primary task, referred in other multitasking literature as a memory for goals \cite{altmann2002memory} or a memory for the problem state \cite{borst_problem_2010}. Previous interruptions literature has shown that interruptions are less disruptive if they are relevant to the primary task  \cite{czerwinski_instant_2000,gould_what_2013, ardissono2009context} as users can better maintain and access memories for the primary task as the interruption would reinforce that memory. Taxing and long interruptions have also been shown to be more disruptive than quicker or simpler disruptions \cite{mark_pace_2012,monk2008effect}. The interruptions used in this work were not likely to have been taxing or long however, as they are reflective of the short and focused information searches identified in the IPA literature \cite{luger_like_2016,dubiel_survey_2018,porcheron_voice_2018} Like in other forms of multitasking, relevance may also be a critical component of the level of interruption experienced with IPAs. Highly relevant IPA tasks may mitigate some of the disrupting effect on primary task performance. In contrast, research on long interruptions \citet{borst_problem_2010} suggests that requests that lead the IPA to give longer or more detailed responses or that include multiple turns may be more disruptive. Further research should focus on these dimensions, exploring how they impact the findings seen in this work.  
 
 \subsection{Interruption Timing in IPA Multitasking}
 Another dimension to consider in future IPA interruptions research is the timing of interruption. Interruptions tend to be less disruptive at subtask boundaries - natural break points within a task \cite{iqbal_disruption_2007,janssen_strategic_2010}. In our research, interruptions were randomly determined and controlled by the experimenter, similar to previous interruptions work \cite{iqbal_leveraging_2006}, but it may be that people may choose to self interrupt with requests to IPAs at natural breakpoints within their tasks. This may lead to less disruption as people vary their interruption lags to encode information about their primary task \cite{trafton_preparing_2003}. Future work may explore the roles of breakpoint timing in IPA multitasking to better understand the role of timing in these interactions.

\subsection{System Generated Interruptions for IPAs}
As IPAs gain the functionality to initiate an interaction, it is also important for developers to know how and when to formulate those initiations. This work shows that some language tasks, particularly tasks involving language production like composing a message or speaking, could be particularly negatively impacted by speech interruptions. On the other hand, language tasks that don’t include production, like reading or transcribing, may not see such negative effects. Likewise, this study shows that tasks which are more cognitively demanding are impacted more negatively by speech interruptions than less cognitively demanding tasks, just as is that case with non-speech HCI interruptions \cite{borst_problem_2010, iqbal2010cars}. 

 Although there is a debate as to the nature and potential for true conversation to occur with an agent \cite{porcheron_voice_2018,clark2019makes}, when designing notifications, interruptions or strategies to conversationally ‘take the floor’ for such agents, it is important to identify when and how to interrupt users through using speech. This is particularly important when users are busy conducting other tasks, like the hands-busy, eyes-busy tasks that would motivate the use of IPAs over other interfaces. As more conversational and system initiated IPA interactions are developed, future work needs to identify strategies for interrupting users in order to start conversations, as well as how these strategies affect performance of a user’s existing tasks. Our research suggests that IPAs may need to consider the potential load level and types of resources that the user is currently utilising when devising an appropriate interruption strategy.  

\section{CONCLUSION}
Our findings indicate that interruptions involving IPA interactions were more disruptive to language generating tasks than to copying tasks. These findings serve as a first step in connecting the well-studied field of interruptions and multitasking with the growing research area on IPAs. This has the benefit of connecting IPA interaction research to cognitive perspectives and theories (e.g. multiple resources theory and working memory) that have underpinned multitasking literature. Bringing IPA interactions into conversation with those theoretical understandings of human cognition may further enrich the study of such interactions. Considering that multitasking is a major perceived benefit of IPA use, it is integral that more research be conducted to unite these fields. Our research suggests that, rather than being beneficial, IPA use may significant negatively impact performance of primary tasks. This is especially true when similar resources are being used by both primary and secondary tasks, such as when generating language. 

\begin{acks}
This research was supported by Science Foundation Ireland (SFI) award number: 13/RC/2106 ADAPT
\end{acks}

\bibliographystyle{ACM-Reference-Format}
\bibliography{multitasking}


\begin{thebibliography}{50}


\ifx \showCODEN    \undefined \def \showCODEN     #1{\unskip}     \fi
\ifx \showDOI      \undefined \def \showDOI       #1{#1}\fi
\ifx \showISBNx    \undefined \def \showISBNx     #1{\unskip}     \fi
\ifx \showISBNxiii \undefined \def \showISBNxiii  #1{\unskip}     \fi
\ifx \showISSN     \undefined \def \showISSN      #1{\unskip}     \fi
\ifx \showLCCN     \undefined \def \showLCCN      #1{\unskip}     \fi
\ifx \shownote     \undefined \def \shownote      #1{#1}          \fi
\ifx \showarticletitle \undefined \def \showarticletitle #1{#1}   \fi
\ifx \showURL      \undefined \def \showURL       {\relax}        \fi
\providecommand\bibfield[2]{#2}
\providecommand\bibinfo[2]{#2}
\providecommand\natexlab[1]{#1}
\providecommand\showeprint[2][]{arXiv:#2}

\bibitem[\protect\citeauthoryear{Alsius, Navarra, and Soto-Faraco}{Alsius
  et~al\mbox{.}}{2007}]%
        {Alsius2007}
\bibfield{author}{\bibinfo{person}{Agn{\`e}s Alsius}, \bibinfo{person}{Jordi
  Navarra}, {and} \bibinfo{person}{Salvador Soto-Faraco}.}
  \bibinfo{year}{2007}\natexlab{}.
\newblock \showarticletitle{Attention to touch weakens audiovisual speech
  integration}.
\newblock \bibinfo{journal}{\emph{Experimental Brain Research}}
  \bibinfo{volume}{183}, \bibinfo{number}{3} (\bibinfo{date}{01 Nov}
  \bibinfo{year}{2007}), \bibinfo{pages}{399--404}.
\newblock
\showISSN{1432-1106}
\urldef\tempurl%
\url{https://doi.org/10.1007/s00221-007-1110-1}
\showDOI{\tempurl}


\bibitem[\protect\citeauthoryear{Altmann and Trafton}{Altmann and
  Trafton}{2002}]%
        {altmann2002memory}
\bibfield{author}{\bibinfo{person}{Erik~M Altmann} {and}
  \bibinfo{person}{J~Gregory Trafton}.} \bibinfo{year}{2002}\natexlab{}.
\newblock \showarticletitle{Memory for goals: An activation-based model}.
\newblock \bibinfo{journal}{\emph{Cognitive science}} \bibinfo{volume}{26},
  \bibinfo{number}{1} (\bibinfo{year}{2002}), \bibinfo{pages}{39--83}.
\newblock


\bibitem[\protect\citeauthoryear{Ardissono, Bosio, Goy, and Petrone}{Ardissono
  et~al\mbox{.}}{2009}]%
        {ardissono2009context}
\bibfield{author}{\bibinfo{person}{Liliana Ardissono}, \bibinfo{person}{Gianni
  Bosio}, \bibinfo{person}{Annamaria Goy}, {and} \bibinfo{person}{Giovanna
  Petrone}.} \bibinfo{year}{2009}\natexlab{}.
\newblock \showarticletitle{Context-aware notification management in an
  integrated collaborative environment}. In \bibinfo{booktitle}{\emph{UMAP 2009
  workshop" Adaptation and Personalization for Web2. 0"}},
  Vol.~\bibinfo{volume}{485}. CEUR, \bibinfo{pages}{21--30}.
\newblock


\bibitem[\protect\citeauthoryear{Aylett, Kristensson, Whittaker, and
  Vazquez-Alvarez}{Aylett et~al\mbox{.}}{2014}]%
        {aylett_none_2014}
\bibfield{author}{\bibinfo{person}{Matthew~P. Aylett}, \bibinfo{person}{Per~Ola
  Kristensson}, \bibinfo{person}{Steve Whittaker}, {and}
  \bibinfo{person}{Yolanda Vazquez-Alvarez}.} \bibinfo{year}{2014}\natexlab{}.
\newblock \showarticletitle{None of a {CHInd}: relationship counselling for
  {HCI} and speech technology}. In \bibinfo{booktitle}{\emph{Proceedings of the
  extended abstracts of the 32nd annual {ACM} conference on {Human} factors in
  computing systems - {CHI} {EA} '14}}. \bibinfo{publisher}{ACM Press},
  \bibinfo{address}{Toronto, Ontario, Canada}, \bibinfo{pages}{749--760}.
\newblock
\showISBNx{978-1-4503-2474-8}
\urldef\tempurl%
\url{https://doi.org/10.1145/2559206.2578868}
\showDOI{\tempurl}


\bibitem[\protect\citeauthoryear{Baayen, Davidson, and Bates}{Baayen
  et~al\mbox{.}}{2008}]%
        {baayen2008mixed}
\bibfield{author}{\bibinfo{person}{R~Harald Baayen}, \bibinfo{person}{Douglas~J
  Davidson}, {and} \bibinfo{person}{Douglas~M Bates}.}
  \bibinfo{year}{2008}\natexlab{}.
\newblock \showarticletitle{Mixed-effects modeling with crossed random effects
  for subjects and items}.
\newblock \bibinfo{journal}{\emph{Journal of memory and language}}
  \bibinfo{volume}{59}, \bibinfo{number}{4} (\bibinfo{year}{2008}),
  \bibinfo{pages}{390--412}.
\newblock


\bibitem[\protect\citeauthoryear{Baayen and Milin}{Baayen and Milin}{2010}]%
        {baayen2010analyzing}
\bibfield{author}{\bibinfo{person}{R~Harald Baayen} {and}
  \bibinfo{person}{Petar Milin}.} \bibinfo{year}{2010}\natexlab{}.
\newblock \showarticletitle{Analyzing reaction times}.
\newblock \bibinfo{journal}{\emph{International Journal of Psychological
  Research}} \bibinfo{volume}{3}, \bibinfo{number}{2} (\bibinfo{year}{2010}),
  \bibinfo{pages}{12--28}.
\newblock


\bibitem[\protect\citeauthoryear{Baddeley and Hitch}{Baddeley and
  Hitch}{1974}]%
        {baddeley1974working}
\bibfield{author}{\bibinfo{person}{Alan~D Baddeley} {and}
  \bibinfo{person}{Graham Hitch}.} \bibinfo{year}{1974}\natexlab{}.
\newblock \showarticletitle{Working memory}.
\newblock In \bibinfo{booktitle}{\emph{Psychology of learning and motivation}}.
  Vol.~\bibinfo{volume}{8}. \bibinfo{publisher}{Elsevier},
  \bibinfo{pages}{47--89}.
\newblock


\bibitem[\protect\citeauthoryear{Bailey, Konstan, and Carlis}{Bailey
  et~al\mbox{.}}{2001}]%
        {bailey_effects_2001}
\bibfield{author}{\bibinfo{person}{Brian~P Bailey}, \bibinfo{person}{Joseph~A
  Konstan}, {and} \bibinfo{person}{John~V Carlis}.}
  \bibinfo{year}{2001}\natexlab{}.
\newblock \showarticletitle{The {Effects} of {Interruptions} on {Task}
  {Performance}, {Annoyance}, and {Anxiety} in the {User} {Interface}.}. In
  \bibinfo{booktitle}{\emph{Interact}}, Vol.~\bibinfo{volume}{1}.
  \bibinfo{pages}{593--601}.
\newblock


\bibitem[\protect\citeauthoryear{Barr, Levy, Scheepers, and Tily}{Barr
  et~al\mbox{.}}{2013}]%
        {barr_random_2013}
\bibfield{author}{\bibinfo{person}{Dale~J Barr}, \bibinfo{person}{Roger Levy},
  \bibinfo{person}{Christoph Scheepers}, {and} \bibinfo{person}{Harry~J Tily}.}
  \bibinfo{year}{2013}\natexlab{}.
\newblock \showarticletitle{Random effects structure for confirmatory
  hypothesis testing: {Keep} it maximal}.
\newblock \bibinfo{journal}{\emph{Journal of memory and language}}
  \bibinfo{volume}{68}, \bibinfo{number}{3} (\bibinfo{year}{2013}),
  \bibinfo{pages}{255--278}.
\newblock


\bibitem[\protect\citeauthoryear{Bates, Maechler, Bolker, Walker, Christensen,
  Singmann, Dai, Scheipl, and Grothendieck}{Bates et~al\mbox{.}}{[n. d.]}]%
        {bates2012package}
\bibfield{author}{\bibinfo{person}{Douglas Bates}, \bibinfo{person}{Martin
  Maechler}, \bibinfo{person}{Ben Bolker}, \bibinfo{person}{Steven Walker},
  \bibinfo{person}{Rune Haubo~Bojesen Christensen}, \bibinfo{person}{Henrik
  Singmann}, \bibinfo{person}{Bin Dai}, \bibinfo{person}{Fabian Scheipl}, {and}
  \bibinfo{person}{Gabor Grothendieck}.} \bibinfo{year}{[n. d.]}\natexlab{}.
\newblock \showarticletitle{Package ‘lme4’}.
\newblock  (\bibinfo{year}{[n. d.]}).
\newblock


\bibitem[\protect\citeauthoryear{Borst, Taatgen, and van Rijn}{Borst
  et~al\mbox{.}}{2010}]%
        {borst_problem_2010}
\bibfield{author}{\bibinfo{person}{Jelmer~P. Borst}, \bibinfo{person}{Niels~A.
  Taatgen}, {and} \bibinfo{person}{Hedderik van Rijn}.}
  \bibinfo{year}{2010}\natexlab{}.
\newblock \showarticletitle{The problem state: {A} cognitive bottleneck in
  multitasking.}
\newblock \bibinfo{journal}{\emph{Journal of Experimental Psychology: Learning,
  Memory, and Cognition}} \bibinfo{volume}{36}, \bibinfo{number}{2}
  (\bibinfo{year}{2010}), \bibinfo{pages}{363--382}.
\newblock
\showISSN{1939-1285, 0278-7393}
\urldef\tempurl%
\url{https://doi.org/10.1037/a0018106}
\showDOI{\tempurl}


\bibitem[\protect\citeauthoryear{Brumby, Cox, Back, and Gould}{Brumby
  et~al\mbox{.}}{2013}]%
        {brumby2013recovering}
\bibfield{author}{\bibinfo{person}{Duncan~P Brumby}, \bibinfo{person}{Anna~L
  Cox}, \bibinfo{person}{Jonathan Back}, {and} \bibinfo{person}{Sandy~JJ
  Gould}.} \bibinfo{year}{2013}\natexlab{}.
\newblock \showarticletitle{Recovering from an interruption: Investigating
  speed- accuracy trade-offs in task resumption behavior.}
\newblock \bibinfo{journal}{\emph{Journal of Experimental Psychology: Applied}}
  \bibinfo{volume}{19}, \bibinfo{number}{2} (\bibinfo{year}{2013}),
  \bibinfo{pages}{95}.
\newblock


\bibitem[\protect\citeauthoryear{Clark, Doyle, Garaialde, Gilmartin, Schlögl,
  Edlund, Aylett, Cabral, Munteanu, and Cowan}{Clark et~al\mbox{.}}{2018}]%
        {clark_state_2018}
\bibfield{author}{\bibinfo{person}{Leigh Clark}, \bibinfo{person}{Phillip
  Doyle}, \bibinfo{person}{Diego Garaialde}, \bibinfo{person}{Emer Gilmartin},
  \bibinfo{person}{Stephan Schlögl}, \bibinfo{person}{Jens Edlund},
  \bibinfo{person}{Matthew Aylett}, \bibinfo{person}{João Cabral},
  \bibinfo{person}{Cosmin Munteanu}, {and} \bibinfo{person}{Benjamin~R.
  Cowan}.} \bibinfo{year}{2018}\natexlab{}.
\newblock \showarticletitle{The {State} of {Speech} in {HCI}: {Trends},
  {Themes} and {Challenges}}.
\newblock \bibinfo{journal}{\emph{Unpublished}} (\bibinfo{year}{2018}).
\newblock
\urldef\tempurl%
\url{https://doi.org/10.13140/rg.2.2.17331.07202}
\showDOI{\tempurl}


\bibitem[\protect\citeauthoryear{Clark, Pantidi, Cooney, Doyle, Garaialde,
  Edwards, Spillane, Murad, Munteanu, Wade, et~al\mbox{.}}{Clark
  et~al\mbox{.}}{2019}]%
        {clark2019makes}
\bibfield{author}{\bibinfo{person}{Leigh Clark}, \bibinfo{person}{Nadia
  Pantidi}, \bibinfo{person}{Orla Cooney}, \bibinfo{person}{Philip Doyle},
  \bibinfo{person}{Diego Garaialde}, \bibinfo{person}{Justin Edwards},
  \bibinfo{person}{Brendan Spillane}, \bibinfo{person}{Christine Murad},
  \bibinfo{person}{Cosmin Munteanu}, \bibinfo{person}{Vincent Wade},
  {et~al\mbox{.}}} \bibinfo{year}{2019}\natexlab{}.
\newblock \showarticletitle{What Makes a Good Conversation? Challenges in
  Designing Truly Conversational Agents}.
\newblock \bibinfo{journal}{\emph{arXiv preprint arXiv:1901.06525}}
  (\bibinfo{year}{2019}).
\newblock


\bibitem[\protect\citeauthoryear{Cockburn and Siresena}{Cockburn and
  Siresena}{2003}]%
        {cockburn2003evaluating}
\bibfield{author}{\bibinfo{person}{Andy Cockburn} {and} \bibinfo{person}{Amal
  Siresena}.} \bibinfo{year}{2003}\natexlab{}.
\newblock \showarticletitle{Evaluating mobile text entry with the Fastap
  keypad}.
\newblock  (\bibinfo{year}{2003}).
\newblock


\bibitem[\protect\citeauthoryear{Cowan, Pantidi, Coyle, Morrissey, Clarke,
  Al-Shehri, Earley, and Bandeira}{Cowan et~al\mbox{.}}{2017}]%
        {cowan_what_2017}
\bibfield{author}{\bibinfo{person}{Benjamin~R Cowan}, \bibinfo{person}{Nadia
  Pantidi}, \bibinfo{person}{David Coyle}, \bibinfo{person}{Kellie Morrissey},
  \bibinfo{person}{Peter Clarke}, \bibinfo{person}{Sara Al-Shehri},
  \bibinfo{person}{David Earley}, {and} \bibinfo{person}{Natasha Bandeira}.}
  \bibinfo{year}{2017}\natexlab{}.
\newblock \showarticletitle{What can i help you with?: infrequent users'
  experiences of intelligent personal assistants}. In
  \bibinfo{booktitle}{\emph{Proceedings of the 19th {International}
  {Conference} on {Human}-{Computer} {Interaction} with {Mobile} {Devices} and
  {Services}}}. \bibinfo{publisher}{ACM}, \bibinfo{pages}{43}.
\newblock


\bibitem[\protect\citeauthoryear{Czerwinski, Cutrell, and Horvitz}{Czerwinski
  et~al\mbox{.}}{2000}]%
        {czerwinski_instant_2000}
\bibfield{author}{\bibinfo{person}{Mary Czerwinski}, \bibinfo{person}{Edward
  Cutrell}, {and} \bibinfo{person}{Eric Horvitz}.}
  \bibinfo{year}{2000}\natexlab{}.
\newblock \showarticletitle{Instant messaging and interruption: {Influence} of
  task type on performance}. In \bibinfo{booktitle}{\emph{{OZCHI} 2000
  conference proceedings}}, Vol.~\bibinfo{volume}{356}.
  \bibinfo{pages}{361--367}.
\newblock


\bibitem[\protect\citeauthoryear{Desjardins and Doherty}{Desjardins and
  Doherty}{2014}]%
        {desjardins2014effect}
\bibfield{author}{\bibinfo{person}{Jamie~L Desjardins} {and}
  \bibinfo{person}{Karen~A Doherty}.} \bibinfo{year}{2014}\natexlab{}.
\newblock \showarticletitle{The effect of hearing aid noise reduction on
  listening effort in hearing-impaired adults}.
\newblock \bibinfo{journal}{\emph{Ear and Hearing}} \bibinfo{volume}{35},
  \bibinfo{number}{6} (\bibinfo{year}{2014}), \bibinfo{pages}{600--610}.
\newblock


\bibitem[\protect\citeauthoryear{Dubiel, Halvey, and Azzopardi}{Dubiel
  et~al\mbox{.}}{2018}]%
        {dubiel_survey_2018}
\bibfield{author}{\bibinfo{person}{Mateusz Dubiel}, \bibinfo{person}{Martin
  Halvey}, {and} \bibinfo{person}{Leif Azzopardi}.}
  \bibinfo{year}{2018}\natexlab{}.
\newblock \showarticletitle{A {Survey} {Investigating} {Usage} of {Virtual}
  {Personal} {Assistants}}.
\newblock \bibinfo{journal}{\emph{arXiv preprint arXiv:1807.04606}}
  (\bibinfo{year}{2018}).
\newblock


\bibitem[\protect\citeauthoryear{Expres}{Expres}{2017}]%
        {listeningexpress}
\bibfield{author}{\bibinfo{person}{Listening Expres}.}
  \bibinfo{year}{2017}\natexlab{}.
\newblock \bibinfo{title}{39 {Nothing} to {Worry} {About}}.
\newblock
\newblock
\urldef\tempurl%
\url{http://www.listeningexpress.com/nce-a/book3/39-Nothing-to-Worry-About.html}
\showURL{%
\tempurl}
\newblock
\shownote{Accessed on 05.07.2018.}


\bibitem[\protect\citeauthoryear{Foroughi, Werner, Barragán, and
  Boehm-Davis}{Foroughi et~al\mbox{.}}{2015}]%
        {foroughi_interruptions_2015}
\bibfield{author}{\bibinfo{person}{Cyrus~K Foroughi}, \bibinfo{person}{Nicole~E
  Werner}, \bibinfo{person}{Daniela Barragán}, {and}
  \bibinfo{person}{Deborah~A Boehm-Davis}.} \bibinfo{year}{2015}\natexlab{}.
\newblock \showarticletitle{Interruptions disrupt reading comprehension.}
\newblock \bibinfo{journal}{\emph{Journal of Experimental Psychology: General}}
  \bibinfo{volume}{144}, \bibinfo{number}{3} (\bibinfo{year}{2015}),
  \bibinfo{pages}{704}.
\newblock


\bibitem[\protect\citeauthoryear{Fraser, Gagn{\'e}, Alepins, and Dubois}{Fraser
  et~al\mbox{.}}{2010}]%
        {fraser2010evaluating}
\bibfield{author}{\bibinfo{person}{Sarah Fraser}, \bibinfo{person}{Jean-Pierre
  Gagn{\'e}}, \bibinfo{person}{Majolaine Alepins}, {and}
  \bibinfo{person}{Pascale Dubois}.} \bibinfo{year}{2010}\natexlab{}.
\newblock \showarticletitle{Evaluating the effort expended to understand speech
  in noise using a dual-task paradigm: The effects of providing visual speech
  cues}.
\newblock \bibinfo{journal}{\emph{Journal of speech, language, and hearing
  research}} \bibinfo{volume}{53}, \bibinfo{number}{1} (\bibinfo{year}{2010}),
  \bibinfo{pages}{18--33}.
\newblock


\bibitem[\protect\citeauthoryear{Galy, Cariou, and Mélan}{Galy
  et~al\mbox{.}}{2012}]%
        {GALY2012269}
\bibfield{author}{\bibinfo{person}{Edith Galy}, \bibinfo{person}{Magali
  Cariou}, {and} \bibinfo{person}{Claudine Mélan}.}
  \bibinfo{year}{2012}\natexlab{}.
\newblock \showarticletitle{What is the relationship between mental workload
  factors and cognitive load types?}
\newblock \bibinfo{journal}{\emph{International Journal of Psychophysiology}}
  \bibinfo{volume}{83}, \bibinfo{number}{3} (\bibinfo{year}{2012}),
  \bibinfo{pages}{269 -- 275}.
\newblock
\showISSN{0167-8760}
\urldef\tempurl%
\url{https://doi.org/10.1016/j.ijpsycho.2011.09.023}
\showDOI{\tempurl}


\bibitem[\protect\citeauthoryear{Gould, Brumby, and Cox}{Gould
  et~al\mbox{.}}{2013}]%
        {gould_what_2013}
\bibfield{author}{\bibinfo{person}{Sandy J.~J. Gould},
  \bibinfo{person}{Duncan~P. Brumby}, {and} \bibinfo{person}{Anna~L. Cox}.}
  \bibinfo{year}{2013}\natexlab{}.
\newblock \showarticletitle{What does it mean for an interruption to be
  relevant? {An} investigation of relevance as a memory effect}.
\newblock \bibinfo{journal}{\emph{Proceedings of the Human Factors and
  Ergonomics Society Annual Meeting}} \bibinfo{volume}{57}, \bibinfo{number}{1}
  (\bibinfo{date}{Sept.} \bibinfo{year}{2013}), \bibinfo{pages}{149--153}.
\newblock
\showISSN{1541-9312}
\urldef\tempurl%
\url{https://doi.org/10.1177/1541931213571034}
\showDOI{\tempurl}


\bibitem[\protect\citeauthoryear{Govender and King}{Govender and King}{2018}]%
        {govender_measuring_2018}
\bibfield{author}{\bibinfo{person}{Avashna Govender} {and}
  \bibinfo{person}{Simon King}.} \bibinfo{year}{2018}\natexlab{}.
\newblock \showarticletitle{Measuring the {Cognitive} {Load} of {Synthetic}
  {Speech} {Using} a {Dual} {Task} {Paradigm}}. In
  \bibinfo{booktitle}{\emph{Interspeech 2018}}. \bibinfo{publisher}{ISCA},
  \bibinfo{pages}{2843--2847}.
\newblock
\urldef\tempurl%
\url{https://doi.org/10.21437/Interspeech.2018-1199}
\showDOI{\tempurl}


\bibitem[\protect\citeauthoryear{Grabowski, Damasio, and Damasio}{Grabowski
  et~al\mbox{.}}{1998}]%
        {grabowski_premotor_1998}
\bibfield{author}{\bibinfo{person}{Thomas~J. Grabowski}, \bibinfo{person}{Hanna
  Damasio}, {and} \bibinfo{person}{Antonio~R. Damasio}.}
  \bibinfo{year}{1998}\natexlab{}.
\newblock \showarticletitle{Premotor and {Prefrontal} {Correlates} of
  {Category}-{Related} {Lexical} {Retrieval}}.
\newblock \bibinfo{journal}{\emph{NeuroImage}} \bibinfo{volume}{7},
  \bibinfo{number}{3} (\bibinfo{date}{April} \bibinfo{year}{1998}),
  \bibinfo{pages}{232--243}.
\newblock
\showISSN{10538119}
\urldef\tempurl%
\url{https://doi.org/10.1006/nimg.1998.0324}
\showDOI{\tempurl}


\bibitem[\protect\citeauthoryear{Hart}{Hart}{2006}]%
        {hart_nasa-task_2006}
\bibfield{author}{\bibinfo{person}{Sandra~G Hart}.}
  \bibinfo{year}{2006}\natexlab{}.
\newblock \showarticletitle{{NASA}-task load index ({NASA}-{TLX}); 20 years
  later}. In \bibinfo{booktitle}{\emph{Proceedings of the human factors and
  ergonomics society annual meeting}}, Vol.~\bibinfo{volume}{50}.
  \bibinfo{publisher}{Sage Publications Sage CA: Los Angeles, CA},
  \bibinfo{pages}{904--908}.
\newblock


\bibitem[\protect\citeauthoryear{Hart and Staveland}{Hart and
  Staveland}{1988}]%
        {hart_development_1988}
\bibfield{author}{\bibinfo{person}{Sandra~G Hart} {and}
  \bibinfo{person}{Lowell~E Staveland}.} \bibinfo{year}{1988}\natexlab{}.
\newblock \showarticletitle{Development of {NASA}-{TLX} ({Task} {Load}
  {Index}): {Results} of empirical and theoretical research}.
\newblock In \bibinfo{booktitle}{\emph{Advances in psychology}}.
  Vol.~\bibinfo{volume}{52}. \bibinfo{publisher}{Elsevier},
  \bibinfo{pages}{139--183}.
\newblock


\bibitem[\protect\citeauthoryear{Ho, Nikolic, Waters, and Sarter}{Ho
  et~al\mbox{.}}{2004}]%
        {ho_not_2004}
\bibfield{author}{\bibinfo{person}{Chih-Yuan Ho}, \bibinfo{person}{Mark~I
  Nikolic}, \bibinfo{person}{Molly~J Waters}, {and} \bibinfo{person}{Nadine~B
  Sarter}.} \bibinfo{year}{2004}\natexlab{}.
\newblock \showarticletitle{Not now! {Supporting} interruption management by
  indicating the modality and urgency of pending tasks}.
\newblock \bibinfo{journal}{\emph{Human Factors}} \bibinfo{volume}{46},
  \bibinfo{number}{3} (\bibinfo{year}{2004}), \bibinfo{pages}{399--409}.
\newblock


\bibitem[\protect\citeauthoryear{Hujiang}{Hujiang}{2017}]%
        {hujiangship}
\bibfield{author}{\bibinfo{person}{Hujiang}.} \bibinfo{year}{2017}\natexlab{}.
\newblock \bibinfo{title}{A lovable eccentric \& {A} lost ship}.
\newblock
\newblock
\urldef\tempurl%
\url{https://st.hujiang.com/topic/15544016853/}
\showURL{%
\tempurl}
\newblock
\shownote{Accessed on 05.07.2018.}


\bibitem[\protect\citeauthoryear{Iqbal and Bailey}{Iqbal and Bailey}{2006}]%
        {iqbal_leveraging_2006}
\bibfield{author}{\bibinfo{person}{Shamsi~T Iqbal} {and}
  \bibinfo{person}{Brian~P Bailey}.} \bibinfo{year}{2006}\natexlab{}.
\newblock \showarticletitle{Leveraging characteristics of task structure to
  predict the cost of interruption}. In \bibinfo{booktitle}{\emph{Proceedings
  of the {SIGCHI} conference on {Human} {Factors} in computing systems}}.
  \bibinfo{publisher}{ACM}, \bibinfo{pages}{741--750}.
\newblock


\bibitem[\protect\citeauthoryear{Iqbal and Horvitz}{Iqbal and Horvitz}{2007}]%
        {iqbal_disruption_2007}
\bibfield{author}{\bibinfo{person}{Shamsi~T Iqbal} {and} \bibinfo{person}{Eric
  Horvitz}.} \bibinfo{year}{2007}\natexlab{}.
\newblock \showarticletitle{Disruption and recovery of computing tasks: field
  study, analysis, and directions}. In \bibinfo{booktitle}{\emph{Proceedings of
  the {SIGCHI} conference on {Human} factors in computing systems}}.
  \bibinfo{publisher}{ACM}, \bibinfo{pages}{677--686}.
\newblock


\bibitem[\protect\citeauthoryear{Iqbal, Ju, and Horvitz}{Iqbal
  et~al\mbox{.}}{2010}]%
        {iqbal2010cars}
\bibfield{author}{\bibinfo{person}{Shamsi~T Iqbal}, \bibinfo{person}{Yun-Cheng
  Ju}, {and} \bibinfo{person}{Eric Horvitz}.} \bibinfo{year}{2010}\natexlab{}.
\newblock \showarticletitle{Cars, calls, and cognition: investigating driving
  and divided attention}. In \bibinfo{booktitle}{\emph{Proceedings of the
  SIGCHI conference on human factors in computing systems}}. ACM,
  \bibinfo{pages}{1281--1290}.
\newblock


\bibitem[\protect\citeauthoryear{Janssen and Brumby}{Janssen and
  Brumby}{2010}]%
        {janssen_strategic_2010}
\bibfield{author}{\bibinfo{person}{Christian~P Janssen} {and}
  \bibinfo{person}{Duncan~P Brumby}.} \bibinfo{year}{2010}\natexlab{}.
\newblock \showarticletitle{Strategic adaptation to performance objectives in a
  dual-task setting}.
\newblock \bibinfo{journal}{\emph{Cognitive science}} \bibinfo{volume}{34},
  \bibinfo{number}{8} (\bibinfo{year}{2010}), \bibinfo{pages}{1548--1560}.
\newblock


\bibitem[\protect\citeauthoryear{Joyen}{Joyen}{2004}]%
        {joyen37}
\bibfield{author}{\bibinfo{person}{Joyen}.} \bibinfo{year}{2004}\natexlab{}.
\newblock \bibinfo{title}{Lesson 37 {The} {Westhaven} {Express}}.
\newblock
\newblock
\urldef\tempurl%
\url{http://www.joyen.net/article/lesson/nce/nce3/200410/258.html}
\showURL{%
\tempurl}
\newblock
\shownote{Accessed on 05.07.2018.}


\bibitem[\protect\citeauthoryear{Luger and Sellen}{Luger and Sellen}{2016}]%
        {luger_like_2016}
\bibfield{author}{\bibinfo{person}{Ewa Luger} {and} \bibinfo{person}{Abigail
  Sellen}.} \bibinfo{year}{2016}\natexlab{}.
\newblock \showarticletitle{Like having a really bad {PA}: the gulf between
  user expectation and experience of conversational agents}. In
  \bibinfo{booktitle}{\emph{Proceedings of the 2016 {CHI} {Conference} on
  {Human} {Factors} in {Computing} {Systems}}}. \bibinfo{publisher}{ACM},
  \bibinfo{pages}{5286--5297}.
\newblock


\bibitem[\protect\citeauthoryear{Mark, Voida, and Cardello}{Mark
  et~al\mbox{.}}{2012}]%
        {mark_pace_2012}
\bibfield{author}{\bibinfo{person}{Gloria Mark}, \bibinfo{person}{Stephen
  Voida}, {and} \bibinfo{person}{Armand Cardello}.}
  \bibinfo{year}{2012}\natexlab{}.
\newblock \showarticletitle{A pace not dictated by electrons: an empirical
  study of work without email}. In \bibinfo{booktitle}{\emph{Proceedings of the
  {SIGCHI} Conference on Human Factors in Computing Systems}}.
  \bibinfo{publisher}{ACM}, \bibinfo{pages}{555--564}.
\newblock


\bibitem[\protect\citeauthoryear{McCutchen}{McCutchen}{1996}]%
        {mccutchen_capacity_1996}
\bibfield{author}{\bibinfo{person}{Deborah McCutchen}.}
  \bibinfo{year}{1996}\natexlab{}.
\newblock \showarticletitle{A capacity theory of writing: {Working} memory in
  composition}.
\newblock \bibinfo{journal}{\emph{Educational Psychology Review}}
  \bibinfo{volume}{8}, \bibinfo{number}{3} (\bibinfo{year}{1996}),
  \bibinfo{pages}{299--325}.
\newblock


\bibitem[\protect\citeauthoryear{Monk, Trafton, and Boehm-Davis}{Monk
  et~al\mbox{.}}{2008}]%
        {monk2008effect}
\bibfield{author}{\bibinfo{person}{Christopher~A Monk},
  \bibinfo{person}{J~Gregory Trafton}, {and} \bibinfo{person}{Deborah~A
  Boehm-Davis}.} \bibinfo{year}{2008}\natexlab{}.
\newblock \showarticletitle{The effect of interruption duration and demand on
  resuming suspended goals.}
\newblock \bibinfo{journal}{\emph{Journal of Experimental Psychology: Applied}}
  \bibinfo{volume}{14}, \bibinfo{number}{4} (\bibinfo{year}{2008}),
  \bibinfo{pages}{299}.
\newblock


\bibitem[\protect\citeauthoryear{Oord, Dieleman, Zen, Simonyan, Vinyals,
  Graves, Kalchbrenner, Senior, and Kavukcuoglu}{Oord et~al\mbox{.}}{2016}]%
        {oord_wavenet:_2016}
\bibfield{author}{\bibinfo{person}{Aaron van~den Oord}, \bibinfo{person}{Sander
  Dieleman}, \bibinfo{person}{Heiga Zen}, \bibinfo{person}{Karen Simonyan},
  \bibinfo{person}{Oriol Vinyals}, \bibinfo{person}{Alex Graves},
  \bibinfo{person}{Nal Kalchbrenner}, \bibinfo{person}{Andrew Senior}, {and}
  \bibinfo{person}{Koray Kavukcuoglu}.} \bibinfo{year}{2016}\natexlab{}.
\newblock \showarticletitle{{WaveNet}: {A} {Generative} {Model} for {Raw}
  {Audio}}.
\newblock \bibinfo{journal}{\emph{arXiv:1609.03499 [cs]}}
  (\bibinfo{date}{Sept.} \bibinfo{year}{2016}).
\newblock
\urldef\tempurl%
\url{http://arxiv.org/abs/1609.03499}
\showURL{%
\tempurl}
\newblock
\shownote{arXiv: 1609.03499.}


\bibitem[\protect\citeauthoryear{Porcheron, Fischer, McGregor, Brown, Luger,
  Candello, and O'Hara}{Porcheron et~al\mbox{.}}{2017}]%
        {porcheron_talking_2017}
\bibfield{author}{\bibinfo{person}{Martin Porcheron}, \bibinfo{person}{Joel~E
  Fischer}, \bibinfo{person}{Moira McGregor}, \bibinfo{person}{Barry Brown},
  \bibinfo{person}{Ewa Luger}, \bibinfo{person}{Heloisa Candello}, {and}
  \bibinfo{person}{Kenton O'Hara}.} \bibinfo{year}{2017}\natexlab{}.
\newblock \showarticletitle{Talking with conversational agents in collaborative
  action}. In \bibinfo{booktitle}{\emph{Companion of the 2017 {ACM}
  {Conference} on {Computer} {Supported} {Cooperative} {Work} and {Social}
  {Computing}}}. \bibinfo{publisher}{ACM}, \bibinfo{pages}{431--436}.
\newblock


\bibitem[\protect\citeauthoryear{Porcheron, Fischer, Reeves, and
  Sharples}{Porcheron et~al\mbox{.}}{2018}]%
        {porcheron_voice_2018}
\bibfield{author}{\bibinfo{person}{Martin Porcheron}, \bibinfo{person}{Joel~E
  Fischer}, \bibinfo{person}{Stuart Reeves}, {and} \bibinfo{person}{Sarah
  Sharples}.} \bibinfo{year}{2018}\natexlab{}.
\newblock \showarticletitle{Voice {Interfaces} in {Everyday} {Life}}. In
  \bibinfo{booktitle}{\emph{Proceedings of the 2018 {CHI} {Conference} on
  {Human} {Factors} in {Computing} {Systems}}}. \bibinfo{publisher}{ACM},
  \bibinfo{pages}{640}.
\newblock


\bibitem[\protect\citeauthoryear{{R Core Team}}{{R Core Team}}{2018}]%
        {rcore}
\bibfield{author}{\bibinfo{person}{{R Core Team}}.}
  \bibinfo{year}{2018}\natexlab{}.
\newblock \bibinfo{booktitle}{\emph{R: A Language and Environment for
  Statistical Computing}}.
\newblock R Foundation for Statistical Computing, Vienna, Austria.
\newblock
\urldef\tempurl%
\url{https://www.R-project.org/}
\showURL{%
\tempurl}


\bibitem[\protect\citeauthoryear{Ratwani, Andrews, Sousk, and Trafton}{Ratwani
  et~al\mbox{.}}{2008}]%
        {ratwani_effect_2008}
\bibfield{author}{\bibinfo{person}{Raj~M Ratwani}, \bibinfo{person}{Alyssa~E
  Andrews}, \bibinfo{person}{Jenny~D Sousk}, {and} \bibinfo{person}{J~Gregory
  Trafton}.} \bibinfo{year}{2008}\natexlab{}.
\newblock \showarticletitle{The effect of interruption modality on primary task
  resumption}. In \bibinfo{booktitle}{\emph{Proceedings of the {Human}
  {Factors} and {Ergonomics} {Society} {Annual} {Meeting}}},
  Vol.~\bibinfo{volume}{52}. \bibinfo{publisher}{Sage Publications Sage CA: Los
  Angeles, CA}, \bibinfo{pages}{393--397}.
\newblock


\bibitem[\protect\citeauthoryear{Salvucci, Taatgen, and Borst}{Salvucci
  et~al\mbox{.}}{2009}]%
        {salvucci2009toward}
\bibfield{author}{\bibinfo{person}{Dario~D Salvucci}, \bibinfo{person}{Niels~A
  Taatgen}, {and} \bibinfo{person}{Jelmer~P Borst}.}
  \bibinfo{year}{2009}\natexlab{}.
\newblock \showarticletitle{Toward a unified theory of the multitasking
  continuum: From concurrent performance to task switching, interruption, and
  resumption}. In \bibinfo{booktitle}{\emph{Proceedings of the SIGCHI
  conference on human factors in computing systems}}. ACM,
  \bibinfo{pages}{1819--1828}.
\newblock


\bibitem[\protect\citeauthoryear{Singmann and Kellen}{Singmann and Kellen}{[n.
  d.]}]%
        {singmann_introduction_nodate}
\bibfield{author}{\bibinfo{person}{Henrik Singmann} {and}
  \bibinfo{person}{David Kellen}.} \bibinfo{year}{[n. d.]}\natexlab{}.
\newblock \showarticletitle{An {Introduction} to {Mixed} {Models} for
  {Experimental} {Psychology}}.
\newblock


\bibitem[\protect\citeauthoryear{Trafton, Altmann, Brock, and Mintz}{Trafton
  et~al\mbox{.}}{2003}]%
        {trafton_preparing_2003}
\bibfield{author}{\bibinfo{person}{J.Gregory Trafton}, \bibinfo{person}{Erik~M
  Altmann}, \bibinfo{person}{Derek~P Brock}, {and} \bibinfo{person}{Farilee~E
  Mintz}.} \bibinfo{year}{2003}\natexlab{}.
\newblock \showarticletitle{Preparing to resume an interrupted task: effects of
  prospective goal encoding and retrospective rehearsal}.
\newblock \bibinfo{journal}{\emph{International Journal of Human-Computer
  Studies}} \bibinfo{volume}{58}, \bibinfo{number}{5} (\bibinfo{date}{May}
  \bibinfo{year}{2003}), \bibinfo{pages}{583--603}.
\newblock
\showISSN{10715819}
\urldef\tempurl%
\url{https://doi.org/10.1016/S1071-5819(03)00023-5}
\showDOI{\tempurl}


\bibitem[\protect\citeauthoryear{Tubbs-Cooley, Cimiotti, Silber, Sloane, and
  Aiken}{Tubbs-Cooley et~al\mbox{.}}{2013}]%
        {tubbs2013observational}
\bibfield{author}{\bibinfo{person}{Heather~L Tubbs-Cooley},
  \bibinfo{person}{Jeannie~P Cimiotti}, \bibinfo{person}{Jeffrey~H Silber},
  \bibinfo{person}{Douglas~M Sloane}, {and} \bibinfo{person}{Linda~H Aiken}.}
  \bibinfo{year}{2013}\natexlab{}.
\newblock \showarticletitle{An observational study of nurse staffing ratios and
  hospital readmission among children admitted for common conditions}.
\newblock \bibinfo{journal}{\emph{BMJ Qual Saf}} \bibinfo{volume}{22},
  \bibinfo{number}{9} (\bibinfo{year}{2013}), \bibinfo{pages}{735--742}.
\newblock


\bibitem[\protect\citeauthoryear{Wickens}{Wickens}{2002}]%
        {wickens_multiple_2002}
\bibfield{author}{\bibinfo{person}{Christopher~D. Wickens}.}
  \bibinfo{year}{2002}\natexlab{}.
\newblock \showarticletitle{Multiple resources and performance prediction}.
\newblock \bibinfo{journal}{\emph{Theoretical Issues in Ergonomics Science}}
  \bibinfo{volume}{3}, \bibinfo{number}{2} (\bibinfo{date}{Jan.}
  \bibinfo{year}{2002}), \bibinfo{pages}{159--177}.
\newblock
\showISSN{1463-922X, 1464-536X}
\urldef\tempurl%
\url{https://doi.org/10.1080/14639220210123806}
\showDOI{\tempurl}


\bibitem[\protect\citeauthoryear{Wiebe, Roberts, and Behrend}{Wiebe
  et~al\mbox{.}}{2010}]%
        {wiebe2010examination}
\bibfield{author}{\bibinfo{person}{Eric~N Wiebe}, \bibinfo{person}{Edward
  Roberts}, {and} \bibinfo{person}{Tara~S Behrend}.}
  \bibinfo{year}{2010}\natexlab{}.
\newblock \showarticletitle{An examination of two mental workload measurement
  approaches to understanding multimedia learning}.
\newblock \bibinfo{journal}{\emph{Computers in Human Behavior}}
  \bibinfo{volume}{26}, \bibinfo{number}{3} (\bibinfo{year}{2010}),
  \bibinfo{pages}{474--481}.
\newblock


\end{thebibliography}

\end{document}